%% file: main.tex
\documentclass[10pt,twocolumn]{article}

\usepackage[T1]{fontenc}
\usepackage[utf8]{inputenc}
\usepackage{lmodern}
\usepackage{microtype}           % better typography

\usepackage[a4paper,margin=1in]{geometry}
\usepackage{mathtools}           % loads amsmath
\usepackage{amssymb,amsthm}
\numberwithin{equation}{section} % equations numbered per section

\usepackage{graphicx}
\graphicspath{{figures/}}        % images in ./figures

\usepackage{booktabs}            % nice rules
\usepackage{array}
\usepackage{makecell}            % header wrapping / multi-line cells
\usepackage[table]{xcolor}       % table shading
\usepackage{siunitx}
\usepackage[labelfont=bf]{caption}
\usepackage{subcaption}

\usepackage{stfloats}            % allow bottom placement of double-column floats
\usepackage{placeins}            % \FloatBarrier
\usepackage{natbib}
\usepackage[hidelinks]{hyperref} % load hyperref near-last

\usepackage{xifthen}             % for conditional figure placeholders
\usepackage[normalem]{ulem}      % \uline etc., keep \emph behavior
\usepackage{supertabular}        % multipage tables in twocolumn

\newcommand{\Stirling}[2]{\genfrac\{\}{0pt}{}{#1}{#2}}
\date{\today}

\title{Uniqueness on a Continuum: Quantifying Tonal Ambiguity Using Information Theory}
\author{Michael Seltenreich$^{1}$\\[4pt]
\small $^{1}$Anthropic PBC\\
\small \texttt{michaelsel@anthropic.com}
}

\begin{document}
    \twocolumn[
        \maketitle
        % NOTE: Material in this optional argument is typeset full-width automatically.
        \begin{abstract}
            We propose a continuous measure of tonal ambiguity that extends the established concept of uniqueness.
            While uniqueness is widely regarded as necessary for tonality, it cannot (i) discriminate among sets that possess it, (ii) capture hierarchical organization in modes of limited transposition, or (iii) account for temporal unfolding.
            To address these limitations, we introduce a companion measure, grounded in information theory, that quantifies tonal ambiguity on a continuous scale.
            The measure applies across pitch-class sets and tuning systems, expanding analytic coverage of tonal relationships and offering a practical tool for theory and analysis.
        \end{abstract}
        \vspace{2em}
    ]

    \section{Introduction}\label{sec:intro}

    Tonality lies at the base of many of the world's musical cultures and is fundamental to Arabic and Turkish maqamat \citep{Farraj2019,Signell2008}, Hindustani and Carnatic ragas \citep{Castellano1984}, and much of Western music \citep{KrumhanslKeil1982}.
    It can be defined as the hierarchical organization of a scale's notes into distinct functional roles, with one note serving as the central “tonic” and another as a contrasting “dominant.”
    \citet{Balzano1982} argues that for such hierarchy to emerge, the set must possess a structural property he calls “uniqueness.”
    This claim was recently supported by \citet{Pelofi2021}, who showed empirically that melodic statistical dependencies are learned more effectively in sets that exhibit uniqueness\citep[see also][]{Raccah2025}.

\begin{figure}[htbp]
        \centering
        \includegraphics[width=\linewidth]{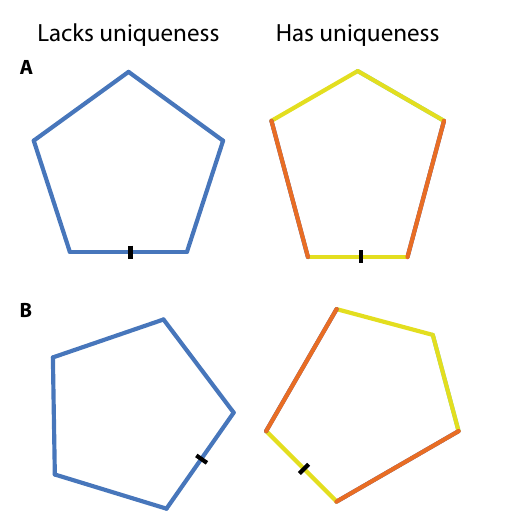}%
        \caption{A) Two pentagons differing in structural properties: the left, with five equal sides, lacks uniqueness; the right, with sides of two distinct lengths (yellow and orange), exhibits uniqueness.
        B) After rotation, only the unique pentagon can be reoriented unambiguously, whereas the uniform pentagon admits multiple indistinguishable orientations.}
        \label{fig:shapes}
    \end{figure}

    A scale exhibits \textit{uniqueness} when relations among its notes permit a single unambiguous transposition.
    This property enables listeners to orient themselves within the conceptual space defined by the scale’s internal relationships.
    \autoref{fig:shapes}A illustrates this with two pentagons. The right pentagon has edges of two distinct lengths (yellow and orange) and exhibits uniqueness; the left, with five equal edges, does not.
    As shown in \autoref{fig:shapes}B, after an unknown rotation the unique pentagon can be restored to its original position with certainty, whereas the uniform pentagon cannot---its identical edges admit multiple indistinguishable orientations.

    The same structural idea carries over to musical scales.
    For example, the major scale exhibits uniqueness, allowing us to identify C–D–E–F–G–A–B unambiguously as the key of C (or one of its modes) rather than as any of its 11 other transpositions; its internal structure makes the orientation clear.
    By contrast, scales that lack uniqueness resist such identification.
    Consider the whole-tone scale: C–D–E–F$\sharp$–G$\sharp$–A$\sharp$ cannot be definitively labeled as a “C whole-tone scale” or a “D whole-tone scale,” since its uniform structure renders every pitch an equally plausible tonic.
    Consequently, the label “C whole-tone scale,” which implies a privileged tonic, is misleading.
    In this respect, the whole-tone scale parallels the uniform pentagon in \autoref{fig:shapes}, as both structures lack uniqueness.
    \citet{Balzano1982} argues that uniqueness is necessary for the emergence of tonality, since the assignment of hierarchical roles depends on unambiguous relations among a scale’s notes.
    Formally this may be so, yet listeners likely tolerate some tonal ambiguity, and a binary framing may obscure more graded interpretations in scales that lack strict uniqueness.
    The modes of limited transposition—among them the whole-tone scale—have long attracted composers precisely because they resist unambiguous transposition.
    Nonetheless, these modes are more structured than atonality: they admit strictly more than one tonal interpretation, but strictly fewer than twelve \citep{Messiaen1944}.
    This allows them to support chromaticism and modulation within clear structural constraints.

    Uniqueness splits scales into two classes—those that have it and those that do not—but offers little guidance about \emph{how strongly} a set points to a tonic.
    Modes of limited transposition, for instance, lack uniqueness by definition, yet they project a distinct musical identity that arbitrary note collections do not: at minimum, they distinguish scale tones from chromatic ones.
    This suggests the need for a continuous measure that captures the degree to which hierarchical relationships can emerge even without uniqueness.

    Even among scales that possess uniqueness, some structures yield a far more salient sense of key than others.
    Moreover, uniqueness concerns a scale as a whole, yet music unfolds sequentially, so uniqueness may not be apparent from the outset, even in scales that formally possess it.

\begin{figure*}[htbp]
        \centering
        \includegraphics[width=0.9\textwidth]{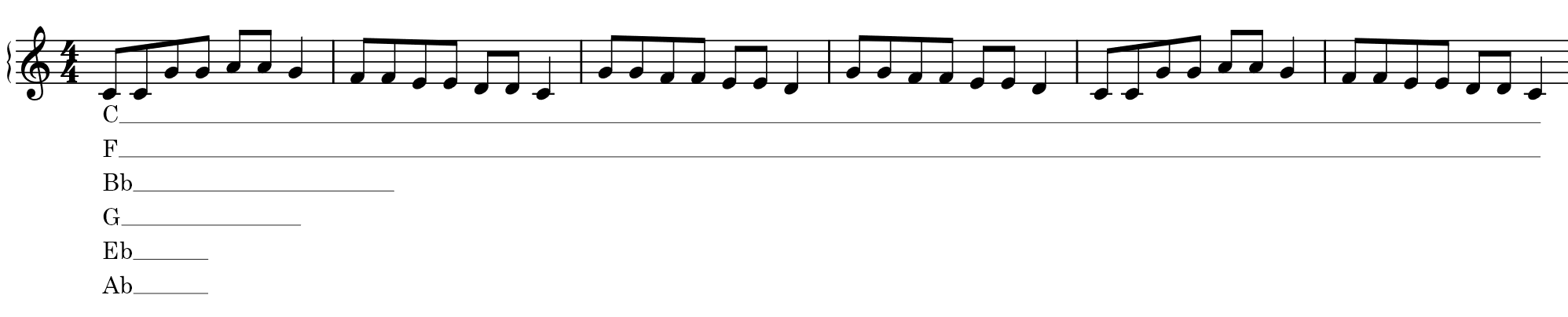}%
        \caption{\emph{Twinkle, Twinkle, Little Star} with possible tonal interpretations marked below.}
        \label{fig:twinkle}
    \end{figure*}

    For instance, although the major scale possesses uniqueness, the tonal center of \emph{Twinkle, Twinkle, Little Star} cannot be determined unambiguously at the outset.
    The first measure (C–G–A–G) is compatible with four major keys (see candidate tonics under the passage in \autoref{fig:twinkle}).
    As the melody unfolds, F$\natural$ excludes G major and E$\natural$ eliminates B$\flat$ major.
    Ultimately, the melody narrows to interpretations in either C or F major, since the absence of a decisive B$\natural$ or B$\flat$ leaves the final tonic (formally) ambiguous.

    Of course, listeners determine a passage’s tonic using many features beyond a tally of pitches.
    For a thorough treatment, see \citet{LerdahlJackendoff1983} and related work \citep{BharuchaKrumhansl1983,Lerdahl2004,Patel1998,Thompson1997}.
    The preceding example simply illustrates the point; it does not suggest that the tonic of \emph{Twinkle, Twinkle, Little Star} is unknowable.
    In practice, listeners often infer underlying scale constituents even when they are not explicitly stated.
    Nonetheless, the passage underscores that uniqueness need not be explicitly expressed for a coherent tonal hierarchy to emerge.

    Importantly, passages like this resemble those belonging to modes of limited transposition in that neither succeeds in structurally identifying a tonic—at least not initially or explicitly—yet both narrow the universe of candidate tonics from twelve to a small subset.

\begin{figure*}[htbp]
        \centering
        \includegraphics[width=0.9\textwidth]{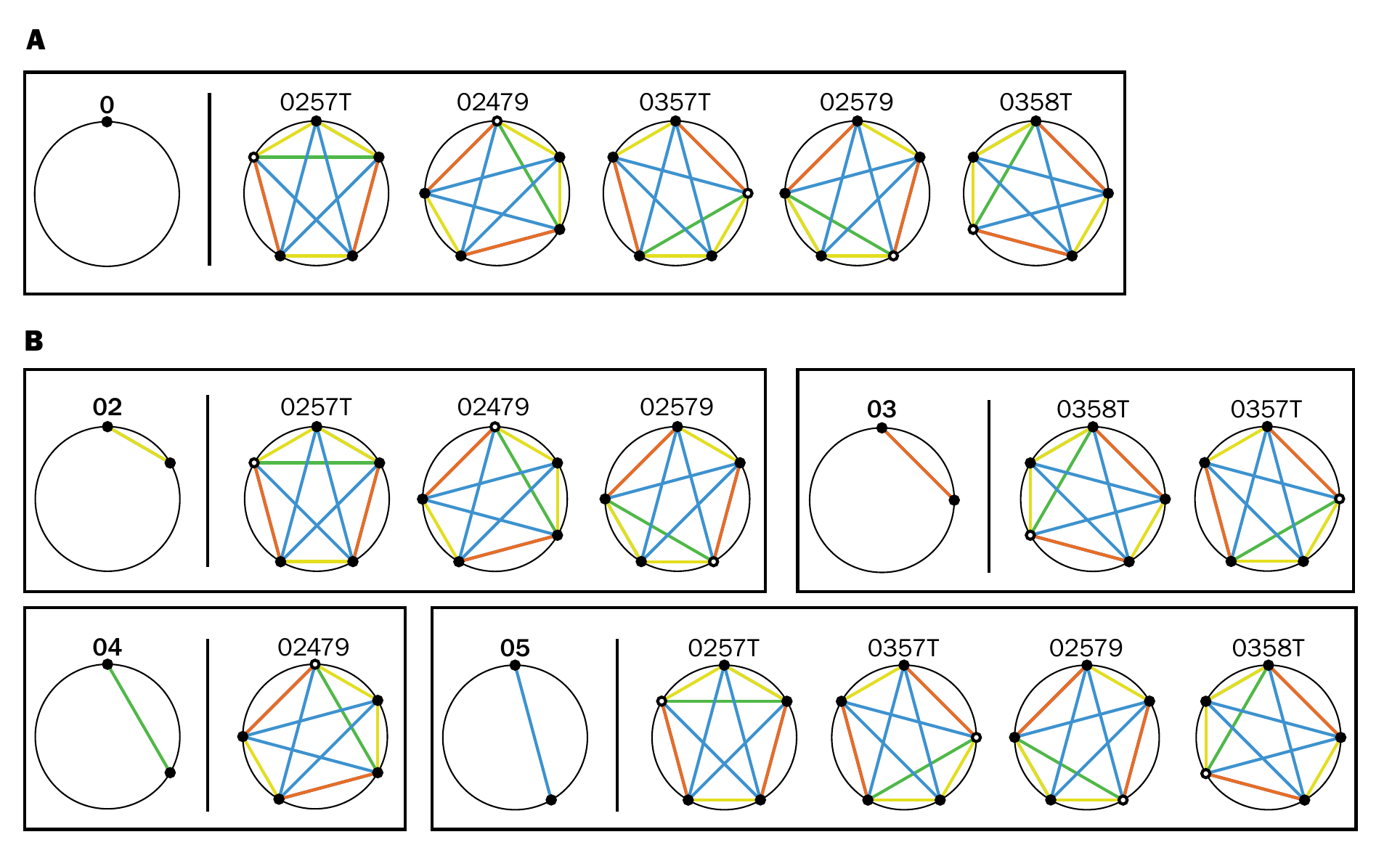}%
        \caption{The anhemitonic pentatonic scale (02479) represented as a pentagon, with edge lengths proportional to composite interval sizes (yellow = 2 semitones, orange = 3, green = 4, blue = 5).
        A white circle marks pitch-class 0 (prime-form orientation).
        \textbf{A)} The five transpositions of 02479 containing pitch-class 0.
        \textbf{B)} Remaining candidate transpositions given intervals 02, 03, 04, and 05, respectively.}
        \label{fig:necklaces}
    \end{figure*}

    In Western practice, tonal ambiguity often lies at the base of musical expression and underpins an important aspect of the art form \citep{NodenSkinner1984, Richards2017, Smith1992, Temperley2007, Uchida1990}.
    To make the challenge concrete, consider the widely prevalent anhemitonic pentatonic set (02479).\footnote{Throughout, we write pitch-class sets as concatenated semitone integers (so $02479 = \{0,2,4,7,9\}$), with $\mathrm{T}=10$ and $\mathrm{E}=11$; pitch class $0$ is taken as C in examples.}
    In this set, a single note reduces the space of possible transpositions from twelve to five, since any pitch may serve as any of the five scale degrees.
    \autoref{fig:necklaces}A illustrates this reduction: when a single pitch is specified (white node), only five transpositions of the set remain (shown in set-theoretical notation and as pentagons; see caption for details).

    While a single note always leaves five possibilities, a second pitch may provide very different amounts of information depending on the interval it forms (\autoref{fig:necklaces}B).
    In some cases (e.g., 04), the specific transposition is determined unambiguously—C and E occur only in \textbf{\uline{C}}–D–\textbf{\uline{E}}–G–A.
    Other intervals (e.g., 05) may leave as many as four of the five candidates; for C and F, the set can be
    B$\flat$–\textbf{\underline{C}}–D–\textbf{\underline{F}}–G,\quad
    E$\flat$–\textbf{\underline{F}}–G–B$\flat$–\textbf{\underline{C}},\quad
    \textbf{\underline{F}}–G–A–\textbf{\underline{C}}–D,\quad
    A$\flat$–B$\flat$–\textbf{\underline{C}}–E$\flat$–\textbf{\underline{F}}.

    These observations suggest the need for a measure that goes beyond binary classification---one that captures degrees of tonal ambiguity in both unique and non-unique scales, and that reflects the sequential unfolding of tonal clarity.
    To this end, we introduce a continuous metric grounded in information theory, which allows us to calculate precisely how much tonal uncertainty remains given an arbitrary collection of pitches.

    Section~\ref{sec:core} presents the core concepts and practical applications of the proposed measure; Section~\ref{sec:math-dev} provides a complete mathematical development, beginning with an accessible introduction to information theory and proceeding to derive the measure across several scale types.

    \section{Core Concepts and Applications}\label{sec:core}

    \noindent\textbf{Who this section is for.}
    Readers who want the \emph{bottom line}: what the paper introduces, how to compute it, and how to use it—without the full pedagogical build-up in \S\ref{sec:math-dev}.
    Readers new to information theory or pitch-class set theory should read \S\ref{sec:math-dev} first and return here as a reference.

    \subsection*{What this paper contributes}
    \begin{enumerate}
        \item A \emph{continuous} measure of tonal ambiguity for any pitch-class collection $S$ in any equal division of the octave (EDO), extending the binary notion of \emph{uniqueness}.
        \item A principled aggregation of information across subset sizes (dyads, trichords, \dots, up to $|S|$) into a single set-level quantity.
        \item A time-aware variant that models melodies as draws \emph{with repetition} and weights results by the expected number of \emph{distinct} pitches.
        \item A cross-set ``disambiguation'' view: how observed notes rule in/out \emph{which} common scale family is likely, not only \emph{which transposition}.
    \end{enumerate}

    \subsection*{Key quantities (use these; derivations in \S\ref{sec:math-dev})}
    Let $c$ be the size of the chromatic space (e.g., $c=12$ in 12-EDO), and let $S\subset\mathbb{Z}_c$ be a pitch-class set of size $m=|S|$. For an observed combination $X\subset\mathbb{Z}_c$ of $k$ distinct pitch classes:
    \begin{itemize}
        \item \textbf{Remaining tonal interpretations within $S$.}
        \[
            t_S(X) \coloneqq \bigl|\{\tau\in \mathbb{Z}_c : X \subseteq \tau{+}S\}\bigr|.
        \]
        This is the number of \emph{candidate transpositions} (and hence candidate tonics) that remain possible for $S$ given $X$. Note that $t_S(X)=1$ identifies the \emph{collection} (e.g., the pitch classes of C major) but not the \emph{mode} (e.g., C Ionian vs.\ A Aeolian); residual modal ambiguity is a separate factor not addressed here. For modes of limited transposition, $t_S(X)$ naturally reflects rotational symmetries.

        \item \textbf{Self-information provided by $X$ (bits).}
        \[
            I_S(X) \coloneqq \log_2\!\left(\frac{c}{t_S(X)}\right).
        \]
        We measure information in bits (base-2 logs); smaller $t_S(X)$ means more information.

        \item \textbf{Cardinality-$k$ expected information (bits).}
        \[
            \mathbb{E}_{\Omega_k(S)}[I]
            \coloneqq \frac{1}{\binom{m}{k}} \sum_{\substack{X\subseteq S\\ |X|=k}} I_S(X).
        \]
        Grouping isomorphic combinations (e.g., by interval class for dyads) yields the same value with combinatorial weights; for dyads these weights are the entries of the interval vector.

        \item \textbf{Set-level measure and its musical reading.}
        Averaging over all non-empty cardinalities,
        \[
            \mathbb{E}_S[I]
            \coloneqq \sum_{k=1}^{m}\!\left(\frac{\binom{m}{k}}{2^m-1}\right)\,\mathbb{E}_{\Omega_k(S)}[I].
        \]
        Report the \emph{average number of tonal interpretations} (geometric-mean sense):
        \[
            \boxed{\;\bar{t}(S)=\dfrac{c}{2^{\mathbb{E}_S[I]}}\;}
        \]
        We refer to $\bar{t}(S)$ as the \emph{Tonal Ambiguity Index} (TAI). By Jensen's inequality, $\bar{t}(S)$ is at most the arithmetic mean of $t_S(X)$ across subsets.
    \end{itemize}

    \paragraph{Time-aware variant (melodic draws).}
    If a passage contains $n$ notes drawn with repetition from $S$ of size $m$, let $P_k$ be the probability that these $n$ draws yield exactly $k$ \emph{distinct} pitch classes:
    \[
        P_k \;=\; m^{-n}\,\Stirling{n}{k}\,\binom{m}{k}\,k!
        \qquad\text{(Eq.~\ref{eq:Pk}).}
    \]
    Then the expected information and its musical reading for an $n$-note passage are
    \[
        \mathbb{E}_S[I \mid n] \;=\; \sum_{k=1}^{m} P_k \,\mathbb{E}_{\Omega_k(S)}[I],
        \qquad
        \bar{t}_n(S) \;=\; \frac{c}{2^{\mathbb{E}_S[I \mid n]}}.
    \]
    In practice, $n{=}8$ offers a good balance: long enough to distinguish 8-note sets (e.g., octatonic) yet short enough to avoid trivial convergence.

    \subsection*{How to compute it (minimal recipe)}
    \begin{enumerate}
        \item \textbf{Fix the universe.} Choose $c$ (e.g., $12$) and a set $S$ (e.g., major, pentatonic, octatonic).
        \item \textbf{Precompute coverage.} For each $\tau\in \mathbb{Z}_c$, form $\tau{+}S$ and record which $X$ (at each $k$) occur.
        \item \textbf{For each $k$.} For all $X{\subseteq}S$ with $|X|{=}k$, compute $t_S(X)$ and $I_S(X)=\log_2(c/t_S(X))$; average to get $\mathbb{E}_{\Omega_k(S)}[I]$.
        \item \textbf{Aggregate.} Combine across cardinalities with the binomial weights to get $\mathbb{E}_S[I]$ and report $\bar{t}(S)=c/2^{\mathbb{E}_S[I]}$.
        \item \textbf{Optional: time-aware.} Weight by $P_k$ to obtain $\bar{t}_n(S)$.
    \end{enumerate}

    \subsection*{Diagnostic combinations and cross-set use}
    \begin{itemize}
        \item \textbf{Diagnostic within a set.} $X$ is \emph{diagnostic} for $S$ if $t_S(X)=1$ (it pins down the tonic/transposition). In the diatonic set, several trichords containing IC\,6 are diagnostic (Table~\ref{tab:trichords}, Fig.~\ref{fig:cadential}).
        \item \textbf{Disambiguating between scale families.} For a family $\mathcal{F}$ (e.g., \{major, harmonic minor, melodic minor, pentatonic, whole-tone, octatonic\}), define
        \[
            t_{\mathcal{F}}(X) \;=\; \bigl|\{\,S\in\mathcal{F}:\exists\,\tau \text{ s.t. } X\subseteq \tau{+}S\,\}\bigr|.
        \]
        The disambiguation gain is $D_{\mathcal{F}}(X)=\log_2(|\mathcal{F}|/t_{\mathcal{F}}(X))$ (bits), defined for combinations $X$ occurring in at least one $S\in\mathcal{F}$: how many families survive after hearing $X$ (applied informally in \S\ref{sec:disambiguating}).
    \end{itemize}

    \subsection*{Headline results (at a glance)}
    Values are $\bar{t}(S)$, the average number of viable tonal interpretations (smaller = less ambiguous).
    \begin{itemize}
        \item \textbf{Common sets with uniqueness.} Major $\approx 2.42$; Pentatonic $\approx 2.29$; Melodic minor $\approx 1.92$; Harmonic minor $\approx 1.87$ (Table~\ref{tab:tonal-interpretations}).
        \item \textbf{Modes of limited transposition.} Whole-tone $\approx 6.00$ (flat across $k$); Octatonic $\approx 4.33$; Augmented $\approx 3.49$ (Table~\ref{tab:tonal-interpretations-mlt}).
        \item \textbf{Temporal convergence (8-note draws).} Major, harmonic minor, and ascending melodic minor approach uniqueness quickly ($\bar{t}_8(S)\!\approx\!1$); pentatonic converges more slowly; whole-tone remains at $6$; octatonic approaches $\sim 4$ (Fig.~\ref{fig:wide}).
    \end{itemize}

    \subsection*{Practical ways to use the metric}
    \begin{enumerate}
        \item \textbf{Analysis.} Given notes $X$ heard in context $S$, report $t_S(X)$ and $I_S(X)$ in bits. Example (diatonic): dyad 01 leaves $t{=}2$ ($\approx 2.58$ bits), while 05 often leaves $t{=}6$ ($1$ bit).
        \item \textbf{Comparing scales.} Prefer smaller $\bar{t}(S)$ for faster \emph{tonal induction}; prefer larger $\bar{t}(S)$ to \emph{sustain} ambiguity.
        \item \textbf{Compositional design.} Use diagnostic combinations ($t{=}1$) to engineer cadential clarity; avoid them (or use high-$t$ combinations) to prolong ambiguity.
        \item \textbf{Pedagogy / cognition.} $\bar{t}_n(S)$ predicts how quickly different sets yield stable tonic inferences as note evidence accrues.
    \end{enumerate}

    \subsection*{Assumptions, knobs, and extensions (brief)}
    \begin{itemize}
        \item The default assumes all $k$-subsets of $S$ are equally likely; style-specific weights (e.g., corpus-derived) can replace the uniform model in $\mathbb{E}_{\Omega_k(S)}[I]$.
        \item Replace $c{=}12$ with any EDO (or any finite chromatic universe) to extend beyond 12-EDO.
        \item The present metric is pitch-class based; timbre, meter, voice-leading, and duration are orthogonal cues you may layer on in other models.
        \item For cross-EDO comparisons, normalize: $\mathrm{NMI}(S) \coloneqq \mathbb{E}_S[I]/\log_2 c$ (higher $=$ less ambiguous) and $\mathrm{NA}(S) \coloneqq 1 - \mathrm{NMI}(S)$ (higher $=$ more ambiguous), so that $\bar{t}(S) = c^{\,\mathrm{NA}(S)}$.
    \end{itemize}

    \section{Mathematical Development}\label{sec:math-dev}

    \textit{The following sections derive the concepts of \S\ref{sec:core} in detail, with a self-contained introduction to information theory for readers less familiar with the framework.}

    \subsection{A ``bit'' about information theory}

    As argued above, the dichotomous framing of uniqueness fails to capture the gradient nature of tonal ambiguity, highlighting the need for a more continuous metric.
    Information theory \citep{Shannon1948} offers a natural framework.
    At its core, the principle is simple: the rarer an event, the more information it provides.

    Consider the word-guessing game hangman, where the goal is to identify a hidden word by guessing one letter at a time.
    Correctly guessing a rare letter such as ``Z'' reduces the pool of candidate words more than guessing a common letter such as ``A,'' since fewer words contain ``Z.''
    As more letters are revealed, the set of possibilities shrinks until only a single word remains.

    The same logic applies to tonality.
    Here, instead of uncovering a hidden word, we seek to identify the tonic of a passage; and instead of adding letters, we add notes.
    Suppose we listen in the context of the major scale.
    Hearing the note F immediately reduces the candidate tonalities from twelve to the seven major keys that contain F (F, G$\flat$, A$\flat$, B$\flat$, C, D$\flat$, and E$\flat$).
    In doing so, it excludes G, A, B, D, and E major.
    Adding a second pitch narrows the field further: for instance, the co-occurrence of F and E restricts the possibilities to just two scales—C major and F major—since these are the only major keys containing both notes.

    This process can be formalized using the concept of \textit{self}\textit{-}\textit{information}, which measures the information gained from a given event.
    It is defined as:

    \begin{align}
        I(E) &= \log_{2} \left( \frac{1}{p(E)} \right), \\
        E &:\ \text{some event}, \\
        p(E) &:\ \text{the probability that this event occurs}.
    \end{align}

    The unit of measure is the \textit{bit}: one bit halves the remaining possibilities.
    If there are initially 100 possible outcomes, one bit reduces the pool to 50; two bits reduce it to 25.

    In Western tonal music, the initial space of possibilities typically consists of the twelve major or minor tonics.
    Acquiring 1 bit of information cuts this space from 12 to 6; 2 bits reduce it to 3.
    To determine the key unambiguously, the space must shrink from 12 to 1, which requires approximately 3.585 bits.
    Put differently, identifying the tonic in Western music entails resolving an initial uncertainty of 3.585 bits.

    \subsection{The diatonic set}

    A single note from the diatonic set reduces the possible transpositions from 12 to 7 (mirroring the pentatonic case of 12 to 5).
    As before, the information contributed by a second note depends on which note it is—but exhaustive listing becomes impractical as set size grows.
    Information theory lets us quantify these contributions systematically.

    In the diatonic set, 2-note combinations vary widely in the information they provide.
    \paragraph{Example.}
    In 12-EDO ($c=12$), the dyad 05 (C--F) leaves six candidate tonics
    (C, D$\flat$, E$\flat$, F, A$\flat$, B$\flat$), so $t_S(X)=6$ and $I_S(X)=\log_2(12/6)=1$ bit.
    By contrast, the dyad 01 leaves $t_S(X)=2$ (e.g., $D\flat$ or $A\flat$ if $0{=}$C), i.e., $\approx 2.58$ bits.

    Formally, the information gained from a given combination can be calculated as\footnote{Sketch: Let $T\in\mathbb{Z}_c$ be the (unknown) tonic with a uniform prior. Given a combination $X$, the surviving tonics form a set of size $t_S(X)$ and $T\mid X$ is uniform on it; hence $H(T)=\log_2 c$, $H(T\mid X)=\log_2 t_S(X)$, and the information gain is $I=H(T)-H(T\mid X)=\log_2(c/t_S(X))$. For non-uniform priors, replace $c$ and $t_S(X)$ by $2^{H(T)}$ and $2^{H(T\mid X)}$.}
    \begin{equation}
        I_S(X) \;=\; \log_{2}\!\left(\frac{c}{t_S(X)}\right).
        \label{eq:derivingSelfInformation}
    \end{equation}
    Unlike the generic self-information $I(E)$ introduced above, $I_S(X)$ measures the per-observation reduction in tonal uncertainty: the information gained about the tonic $T$ from observing $X$ within set $S$.
    We measure information in bits (base-2 logarithms). Put simply, $c$ is the size of the chromatic space (12 in 12-EDO) and $t_S(X)$ is the number of candidate \emph{tonics} (i.e., tonic pitch classes) that remain possible given $X$.

    Thus, 01 yields approximately $2.58$ bits, while 05 yields $1$ bit. Conversely, the number of remaining tonics can be recovered from bits of information as
    \begin{equation}
        t_S(X) \;=\; \frac{c}{2^{\,I_S(X)}}.
        \label{eq:remaining_transpositions}
    \end{equation}

    Table~\ref{tab:diatonic-2note} summarizes the information values for all 2-note combinations in the diatonic set, showing both the number of candidate tonics and their associated probabilities based on the interval vector $\langle 2,5,4,3,6,1\rangle$ (the multiplicity of each interval class 1--6 within the set).

    \input{tables/tab-diatonic-2note}

    We refer to the number of candidate \emph{tonics}—i.e., pitch classes that can serve as the tonic under set $S$ given the observed notes—as the extent of tonal ambiguity.
    However, it is important to recall that uniqueness is a property of the set as a whole, not of specific subsets.
    Thus, while computing tonal ambiguity for particular combinations is useful, it does not by itself capture the overall tonal clarity afforded by the set.
    To do so, we must weigh the high information gained from rare combinations against their low probability of occurring.

    The earlier hangman analogy makes this clear.
    Correctly guessing the letter ``Z'' provides more information than guessing ``A,'' but ``Z'' is also much less likely to be correct in the first place.
    Similarly, in a musical context, highly diagnostic combinations (such as rare intervals) may drastically narrow the set of possibilities, but their infrequent occurrence means they contribute less to the average experience of tonal induction.
    Thus, we need a method that balances the information each combination provides with the likelihood of its appearance.

    For this, we turn to the expected (average) information, following \citet{Shannon1948}.
    Formally:

    \begin{equation}
        \mathbb{E}[I] = \sum_{x} p(x) \cdot I(x)
        \label{eq:expected-info}
    \end{equation}

    where $p(x)$ is the probability of a given interval class (or note combination) $x$ occurring in the (diatonic or other) set, and $I(x)$ is the self-information of that combination.
    Applying this to the diatonic set yields:

    \begin{align}
        \mathbb{E}_{\text{set}~\Omega_{2}}[I] \;\approx\; &
        \underbrace{\left( \frac{2}{21} \cdot 2.58 \right)}_{\substack{\text{01:}\\ p(x)\cdot I(x)}} +
        \left( \frac{5}{21} \cdot 1.26 \right) +
        \left( \frac{4}{21} \cdot 1.58 \right)
        \notag \\[4pt]
        &+ \left( \frac{3}{21} \cdot 2.00 \right) +
        \left( \frac{6}{21} \cdot 1.00 \right)
        \notag \\[4pt]
        &+ \left( \frac{1}{21} \cdot 2.58 \right)
        \;\approx\; 1.54
        \label{eq:expectation-example}
    \end{align}

    This result indicates that, on average, a 2-note combination reduces the space of possible transpositions by a factor of $2^{1.54} \approx 2.91$, leaving about $4$ viable options out of $12$.

    Interestingly, the comparison between the number of instances of each interval and the number of candidate transpositions reveals a striking pattern, consistent with the rare-interval hypothesis \citep{Browne1981,Butler1989}.
    In most cases, these values coincide; the exception is IC6, which occurs once (the dyad 06, i.e., F--B in C major) but still allows two candidate transpositions (C major or G$\flat$ major if 0 = C).
    This discrepancy reflects the special case of symmetrical divisions of the octave (e.g., 06, 048, 0369, 02468T), which yield rotational symmetries and underpin modes of limited transposition.
    Thus, expected information effectively captures the tonal ambiguity arising from such symmetries.

    While this measure provides a richer view than raw 2-note statistics, it remains limited to pairs.
    To better understand tonal ambiguity, we extend the analysis to trichords.
    The same method applies: we calculate the self-information of each trichord, weighted by its probability of occurrence within the set.
    Unlike dyads, however, probabilities cannot be derived directly from the interval vector.
    Instead, we enumerate the trichords present in the set, using the binomial coefficient formula $\binom{n}{k} = \frac{n!}{(n-k)! \, k!}$ to determine the total number of trichords (35 in a 7-note set).

    \input{tables/tab-trichords}

\begin{figure}[htbp]
        \centering
        \includegraphics[width=\linewidth]{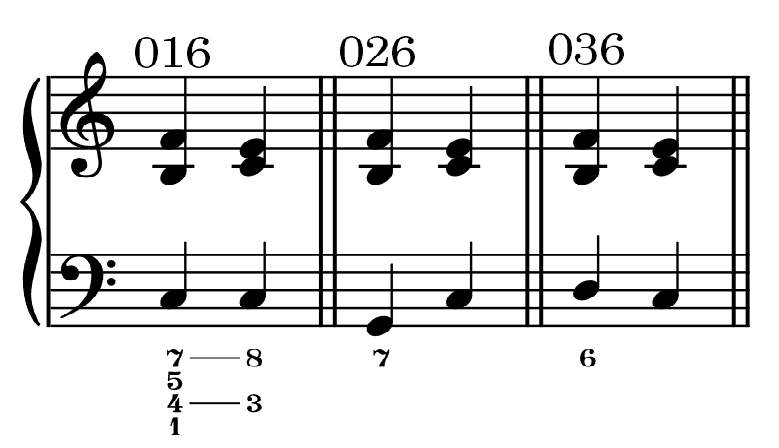}%
        \caption{Diagnostic trichords of the diatonic set in cadential contexts.}
        \label{fig:cadential}
    \end{figure}

    As Table~\ref{tab:trichords} shows, certain trichords alone are sufficient to uniquely identify the scale's transposition.
    We term these \textit{diagnostic combinations}, as they effectively diagnose the set's identity from a subset.
    In the diatonic set, diagnostic trichords include 016, 026, 036, 046, and 056—each sufficient to determine the remaining notes.
    Perhaps unsurprisingly, such combinations frequently appear in cadential contexts (see \autoref{fig:cadential}), with the dominant seventh chord containing diagnostic trichords.
    Indeed, in this set, any 3-note combination containing IC\,6 (a tritone) is a diagnostic combination.

    Although diagnostic trichords yield more information than any dyad, some dyads still outperform certain trichords: intervals 01, 03, 04, and 06 each provide more information than the trichord 027.
    This overlap across cardinalities motivates assessing the full combination space rather than individual cases.

    By calculating expected information across all dyads ($\Omega_2$), we find that a random 2-note combination provides an average of about 1.54 bits of information, leaving roughly 4.1 candidate tonics.
    For trichords ($\Omega_3$), the average rises to about 2.16 bits, leaving about 2.7 candidate tonics.
    Thus, while trichords are generally more informative, the distributions overlap enough to warrant comparison across cardinalities.

    The next step is to synthesize a single value for the set as a whole by computing expected information for each cardinality and weighting by relative frequency.

    Table~\ref{tab:entropy-diatonic} lists the expected-information values for all cardinalities in the diatonic set, derived using binomial coefficients to enumerate the total number of combinations at each size.
    A 7-note set yields 127 possible combinations (7+21+35+35+21+7+1).
    The weighted sum is:

    \begin{align}
        \mathbb{E}[I]_{\text{diatonic}} &= \sum_{k=1}^{7} \frac{\binom{7}{k}}{127} \; \mathbb{E}_{\Omega_k}[I]
        \label{eq:diatonic-avg} \\[4pt]
        \intertext{Substituting the values from Table~\ref{tab:entropy-diatonic}:}
        \mathbb{E}[I]_{\text{diatonic}} \;\approx\; &
        \underbrace{\left( \frac{7}{127} \cdot 0.78 \right)}_{\substack{k=1:\\ \text{weight}\cdot \mathbb{E}_{\Omega_1}}} +
        \underbrace{\left( \frac{21}{127} \cdot 1.54 \right)}_{\substack{k=2:\\ \text{weight}\cdot \mathbb{E}_{\Omega_2}}} 
        \notag \\[4pt]
        &+
        \underbrace{\left( \frac{35}{127} \cdot 2.16 \right)}_{\substack{k=3:\\ \text{weight}\cdot \mathbb{E}_{\Omega_3}}} +
        \left( \frac{35}{127} \cdot 2.61 \right) 
        \notag \\[4pt]
        &+
        \left( \frac{21}{127} \cdot 2.98 \right) +
        \left( \frac{7}{127} \cdot 3.30 \right) 
        \notag \\[4pt]
        &+
        \left( \frac{1}{127} \cdot 3.58 \right)
        \;\approx\; 2.32
        \label{eq:diatonic-sub}
    \end{align}

    \input{tables/tab-entropy-diatonic}

    This corresponds to about $2.42$ candidate tonal interpretations on average.
    For brevity, we call this value the \emph{Tonal Ambiguity Index} (TAI) of the set:
    \[
        \mathrm{TAI}(S) \;=\; \bar{t}(S) \;=\; \frac{c}{2^{\mathbb{E}_S[I]}}.
    \]
    Smaller TAI indicates less ambiguity (fewer viable tonics); larger TAI indicates more ambiguity.

    Since raw values in bits may feel abstract, we recommend reporting ambiguity as the number of possible tonal interpretations (via Eq.~\ref{eq:remaining_transpositions}), which preserves rigor while remaining musically interpretable.

    \subsection{Common scales with uniqueness}

    The TAI enables direct comparison of scales, both at individual cardinalities and at the set level.
    Table~\ref{tab:tonal-interpretations} presents ambiguity values ($t$) for several widely used scales, showing the number of possible tonal interpretations for note combinations ranging from one to seven notes, as well as the overall ambiguity of each set.

    On average, the major scale exhibits the greatest tonal ambiguity, with $\bar{t}\approx 2.42$ possible interpretations.
    This is followed by the pentatonic scale (t=2.29), the melodic minor (t=1.92), and the harmonic minor (t=1.87).
    Put differently, hearing a random combination of notes from the harmonic minor scale is more likely to allow its transposition to be identified unambiguously than a comparable combination drawn from the major scale.

    In practical terms, the major scale's TAI is roughly 30\% higher than the harmonic minor's ($2.42$ vs.\ $1.87$), indicating that the major scale structurally supports more sustained tonal ambiguity---a meaningful dimension of compositional practice.

    \input{tables/tab-tonal-interpretations}

    \subsection{Modes of limited transposition}

    As noted in the introduction, modes of limited transposition were a central motivation for this study.
    Because they do not possess the uniqueness property, they present an ideal testing ground for a measure that quantifies tonal ambiguity beyond a simple binary classification.
    Their enduring popularity may well stem from precisely this balance: they convey a sense of tonality while preserving a structural ambiguity absent in unstructured atonal passages.

    Our measure of ambiguity enables a meaningful comparison among such modes, even in the absence of uniqueness.
    Table~\ref{tab:tonal-interpretations-mlt} presents three common examples: the whole-tone scale (02468T), the octatonic scale (0235689E), and the augmented scale (03478E).
    For each, the table shows the number of tonal interpretations (t) available across different combination cardinalities, as well as the overall ambiguity of the set.

    The results reveal that the whole-tone scale is the most ambiguous, averaging t=6 possible tonal interpretations.
    The octatonic scale follows with $t \approx 4.33$, while the augmented scale exhibits the least ambiguity of the three, with $t \approx 3.49$.
    These findings underscore how, even without uniqueness, different modes of limited transposition afford markedly different capacities for sustaining tonal ambiguity.

    \input{tables/tab-tonal-interpretations-mlt}

    \subsection{Consolidating insights}

    Table~\ref{tab:scale-ambiguity} presents the ambiguity values for the seven common scales examined in this study.
    This measure situates scales with and without uniqueness on a single continuum.
    We can now not only state that the octatonic is more ambiguous than the major scale, but quantify that it has nearly twice the tonal ambiguity---a comparison that uniqueness alone cannot express.

    \input{tables/tab-scale-ambiguity}

    \subsection{Allowing for time}

    The ambiguity values computed thus far rest on an important assumption: that each note drawn from a scale is new, not previously heard.
    This assumption overlooks the temporal nature of music, where notes may repeat before new ones appear.
    To account for this, we can instead imagine constructing melodies by randomly drawing notes from a scale and then tallying the number of \textit{distinct} pitches in the result.
    Rather than averaging across all possible combinations in every cardinality, this approach focuses on the combinations most likely to occur, yielding a more nuanced and musically representative measure of ambiguity.

    For example, the 8-note melody A–C–G–G–E–C–A–E contains four distinct pitches (A, C, E, G), leaving three candidate tonics: C, F, and G major.
    But another 8-note melody might contain seven distinct pitches, or just one.
    To compute expected information, we therefore need the probability of encountering exactly $k$ distinct pitches in $n$ draws.

    The probability of drawing exactly $k$ distinct pitches when selecting $n$ notes from a scale of cardinality $m$ is given by \citet{Feller1968}:

    \begin{equation}
        P_k = m^{-n} \, \Stirling{n}{k} \, \binom{m}{k} \, k!
        \label{eq:Pk}
    \end{equation}

    For example, drawing $n = 8$ notes from the diatonic scale ($m = 7$) produces:

    \begin{align*}
        P_1 &= 7^{-8} \, \Stirling{8}{1} \, \binom{7}{1} \cdot 1! \\
        &= 7^{-8} \cdot 1 \cdot 7 \cdot 1
        \approx 0.000
    \end{align*}

    For two distinct notes:

    \begin{align*}
        P_2 &= 7^{-8} \, \Stirling{8}{2} \, \binom{7}{2} \cdot 2! \\
        &= 7^{-8} \cdot 127 \cdot 21 \cdot 2
        \approx 0.001
    \end{align*}

    And so on:
    \begin{align*}
        P_3 &\approx 0.035, \quad
        P_4 \approx 0.248, \quad
        P_5 \approx 0.459, \\
        P_6 &\approx 0.233, \quad
        P_7 \approx 0.024.
    \end{align*}

    Applying these probabilities as weights to the corresponding expected-information values yields:
    \[
        P_1 \cdot \mathbb{E}_{\Omega_1}[I] \;+\; \cdots \;+\; P_7 \cdot \mathbb{E}_{\Omega_7}[I]
        \;\approx\; 2.95 \ \text{bits}.
    \]

    On average, eight draws from the diatonic scale reduce the transpositional space to about 1.55 candidate tonics.

\begin{figure*}[htbp]
        \centering
        \includegraphics[width=0.9\textwidth]{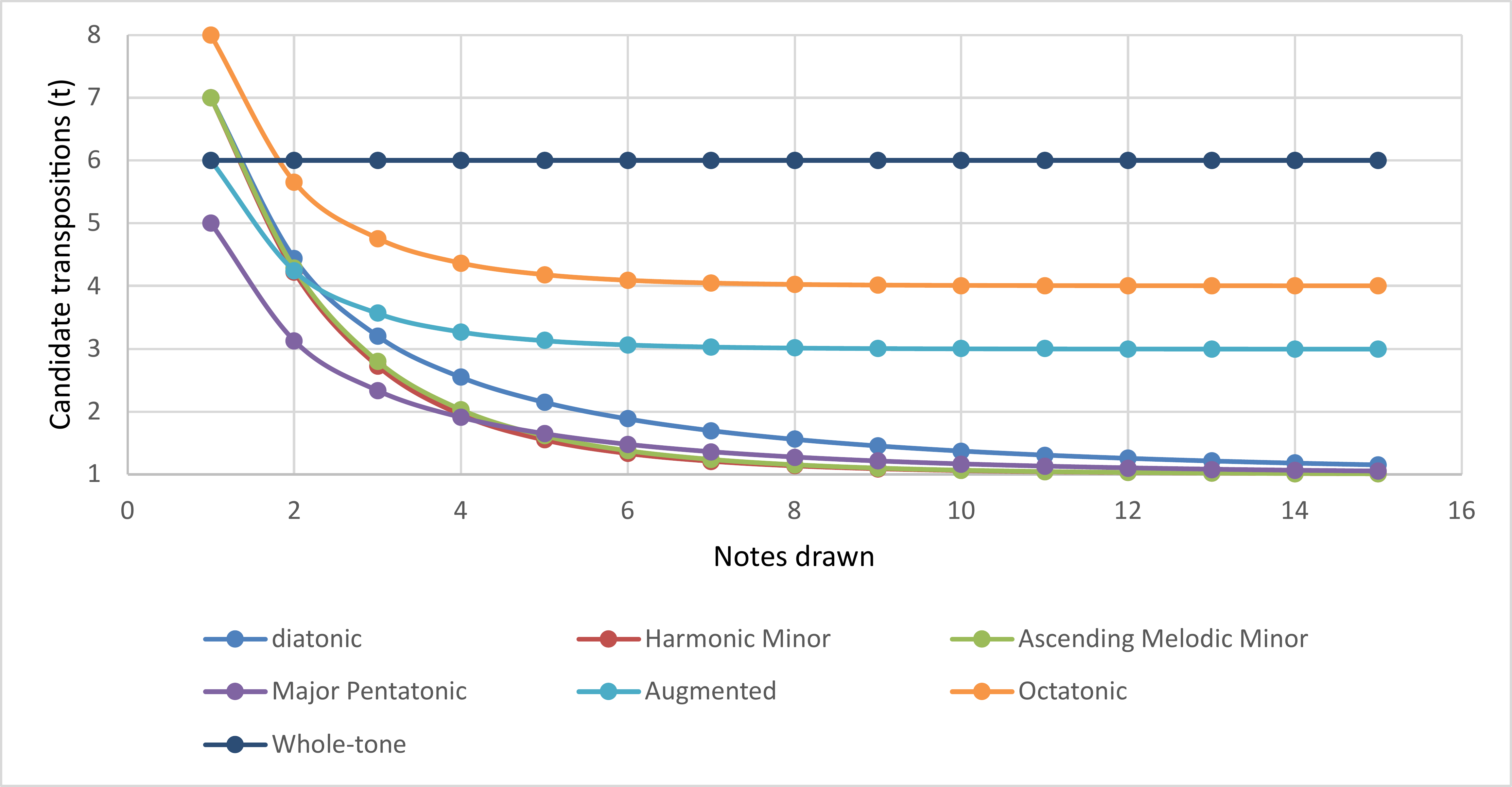}%
        \caption{The number of tonal interpretations available in common scales as a function of the number of notes drawn from them at random.
        The y-axis represents the number of tonal interpretations available.
        The x-axis refers to the number of draws.
        These values were computed using Eq.~\ref{eq:Pk}.}
        \label{fig:wide}
    \end{figure*}

    \autoref{fig:wide} shows how the amount of information gained varies with melody length across several common scales.
    The octatonic scale begins with eight candidate tonalities and converges to four.
    The whole-tone scale remains fixed at six throughout.
    The major, harmonic minor, and ascending melodic minor each begin with seven and converge rapidly to uniqueness (one interpretation).
    The pentatonic scale begins with five and eventually converges to uniqueness as well, though somewhat more slowly than the minor scales.

\begin{figure*}[htbp]
        \centering
        \includegraphics[width=0.9\textwidth]{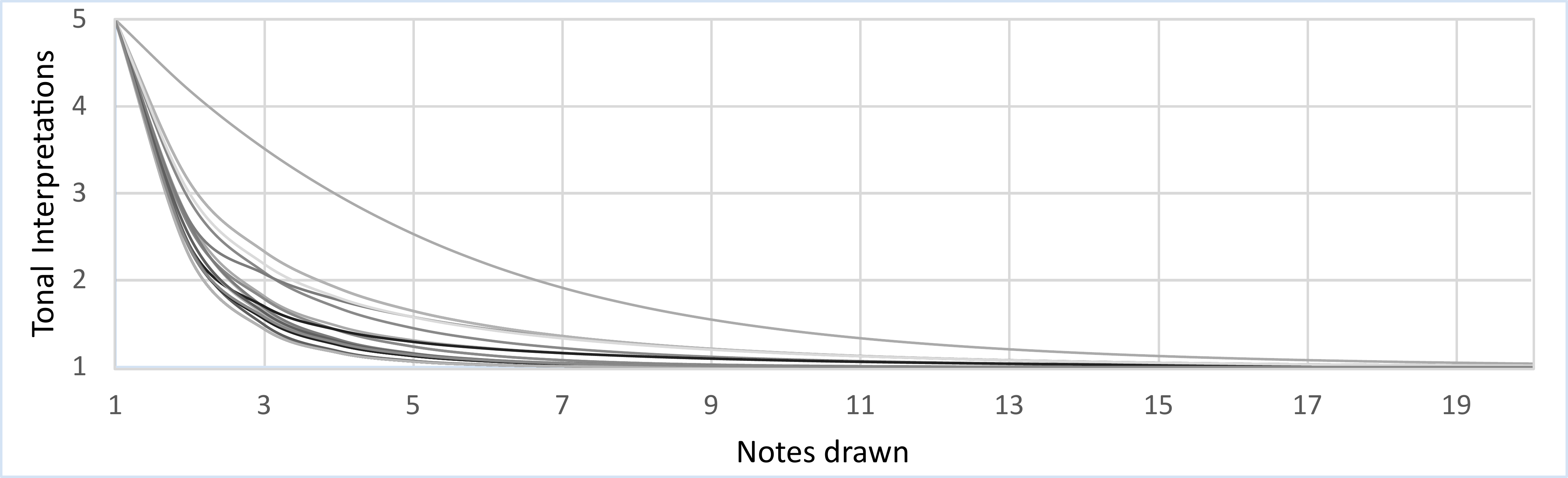}%
        \caption{All 5-note sets in 12-EDO (twelve equal divisions of the octave) and the number of tonal interpretations they leave as a function of the number of notes drawn at random from the set.}
        \label{fig:time}
    \end{figure*}

    But how do we determine the appropriate number of notes to draw to represent a set?
    As \autoref{fig:time} illustrates, the extremes each present problems.
    A single note offers virtually no information about the set, while an infinite sequence of notes drives the measure toward the minimal ambiguity value, obscuring the subtler tonal distinctions.
    For practical purposes, eight notes were chosen as a compromise: enough to allow full representation even in 8-note sets such as the octatonic, yet not so many as to guarantee convergence.
    Still, this remains an arbitrary decision.

    One possible method of summarizing the temporal behavior of a set is to measure the \textit{area under the convergence curve} (AUC).
    In this framing, the slower a scale converges, the larger the enclosed area.
    However, this approach has three notable drawbacks.
    First, computing AUC is more complex than earlier calculations.
    Second, the AUC reflects a mixture of the rate of convergence and the initial degree of ambiguity, making it harder to interpret in isolation.
    For example, the octatonic scale converges quickly to four interpretations ($AUC \approx 0.946$), while the major scale converges more slowly ($AUC \approx 3.803$).
    Third—and most importantly—AUC values are difficult to interpret musically.

    That said, AUC and ambiguity values are highly correlated ($r^2 > 0.98$ for both 5- and 7-note scales), suggesting a strong intrinsic relationship.
    Thus, while AUC may serve as a form of validation, the ambiguity values developed earlier remain preferable: they are rigorous, musically interpretable, and provide a direct means of quantifying tonal ambiguity across different sets.

    \subsection{Disambiguating sets}\label{sec:disambiguating}

    So far we have assumed the underlying set is known and asked only which transposition is in play.
    In practice, listeners must also identify the set itself: not merely that a piece is in C, but whether it is in C major, C minor, or another set entirely \citep{Krumhansl1979}.

    We therefore ask: can an observed combination \textit{disambiguate between different sets}?
    Formally, the answer is ``barely,'' but in practice—since not all sets occur with equal frequency—the answer may well be ``yes.''

    Table~\ref{tab:comb-cardinality} presents the average number of possible 7-note sets remaining after eliminating candidates based on combinations of varying cardinalities.
    Across the 66 distinct 7-note sets in 12-EDO, the results appear daunting: even after hearing five of the seven notes, an average of nearly 20 candidate sets remain.
    The most diagnostic combination, 01369, still leaves 13 possibilities.
    Identifying a specific set among the full field of 66 thus seems highly impractical.

    \input{tables/tab-comb-cardinality}

    However, as previously noted, not all sets are equally likely in Western music.
    In fact, the effective pool of candidates is much smaller, consisting primarily of the diatonic set (major and its modes), the harmonic and melodic minors, the major pentatonic and its modes, the whole-tone scale, and the octatonic scale.
    The precise list may vary slightly, but the principle remains: real-world listening involves a limited set of common tonal frameworks.

    Within this restricted pool, disambiguation becomes far more practical (Table~\ref{tab:avg-viable-sets}): 0145 virtually guarantees the harmonic minor, while 0167 points strongly to the octatonic.

    \input{tables/tab-avg-viable-sets}

    \section{Conclusion}\label{sec:conclusion}

    The property of uniqueness provides complete orientation within a collection of notes, ensuring that its transposition can be identified without ambiguity.
    In scales that fulfill uniqueness, this property supports the emergence of a clear and singular tonality.
    Yet if we allow that tonality can tolerate some degree of uncertainty, uniqueness alone proves too restrictive a condition for the presence of hierarchical relationships among a scale's notes.
    As such, uniqueness cannot capture the gradient nature of tonal clarity across different note collections.

    This paper has proposed a complementary measure that places unique and non-unique scales on a single continuum of tonal ambiguity, offering a more nuanced view of each set's capacity for tonal organization.

    Using information theory, we examined the information gained from combinations of varying cardinalities, aggregated these into a set-level index, and modeled the temporal dimension by treating melodies as draws with repetition.
    We also showed that while subset analysis alone cannot identify a specific set among all 66 heptachords, restricting attention to common Western scales makes disambiguation practical.

    An appendix to this paper lists every Tn-class of one to six notes occurring in the scales discussed, showing both the number of tonal interpretations that remain in the context of each scale and the number of viable candidate sets.
    This comprehensive resource supports further application of the ambiguity measure and demonstrates its utility for music theory and analysis.

% --- References (natbib) ---
    \bibliographystyle{apalike}
    \bibliography{refs}

    \clearpage

% Shading helpers
    \newcommand{\sone}[1]{\cellcolor{gray!20}{#1}} % single interpretation "1"
    \newcommand{\diag}{\rowcolor{green!12}}        % Possible Sets = 1 row highlight

% ---------- Appendix: full table ----------
    \section*{Appendix: Possible Transpositions by Combination}

    % ---- begin squeeze ----
    \begingroup
    \footnotesize                           % smaller font (try \footnotesize if this is too small)
    \setlength{\tabcolsep}{5pt}            % narrower column padding (default is 6pt)
%    \renewcommand{\arraystretch}{0.9}     % tighter rows (1.0 is default)

% (optional) slightly tighter booktabs spacing
    \setlength{\aboverulesep}{0.3ex}
    \setlength{\belowrulesep}{0.3ex}

% Repeated head/foot for supertabular
    \tablefirsthead{%
        \toprule
        \makecell[l]{Combo} &
        \makecell[c]{M} & \makecell[c]{mm} & \makecell[c]{hm} &
        \makecell[c]{WT} & \makecell[c]{O} & \makecell[c]{P} &
        \makecell[c]{A} & \makecell[c]{Possible\\Sets} \\
        \midrule
    }
    \tablehead{%
        \toprule
        \makecell[l]{Combo} &
        \makecell[c]{M} & \makecell[c]{mm} & \makecell[c]{hm} &
        \makecell[c]{WT} & \makecell[c]{O} & \makecell[c]{P} &
        \makecell[c]{A} & \makecell[c]{Possible\\Sets} \\
        \midrule
    }
    \tabletail{%
        \midrule
        \multicolumn{9}{r}{\emph{Continued on next column/page}}\\
    }
    \tablelasttail{\bottomrule}

% your table content:
    \input{tables/tab-appendix}

    \endgroup
% ---- end squeeze ----

    \paragraph{How to reproduce values in appendix}
    \small
    For each set $S\subset\mathbb{Z}_c$ and cardinality $k$:
    (1) enumerate all $k$-subsets $X\subseteq S$ (or group by isomorphism for dyads via the interval vector);
    (2) compute $t_S(X)=|\{\tau : X\subseteq\tau{+}S\}|$;
    (3) set $I_S(X)=\log_2(c/t_S(X))$ and average over $X$ to obtain $\mathbb{E}_{\Omega_k(S)}[I]$;
    (4) aggregate across cardinalities with weights $\binom{|S|}{k}/(2^{|S|}-1)$ to get $\mathbb{E}_S[I]$ and report $\mathrm{TAI}(S)=c/2^{\mathbb{E}_S[I]}$.
    For time-aware results at length $n$, weight the cardinality-specific expectations by $P_k$ from Eq.~\ref{eq:Pk}.

\end{document}

%% file: tables/tab-diatonic-2note.tex
\begin{table}[htbp]
\centering
\caption{Information-theoretic values for 2-note combinations in the diatonic set.
Columns list: pitch class combination, number of candidate transpositions ($t$),
information gain in bits ($I_S(X)$), number of instances in the scale, and the probability
of each combination occurring $p(x)$.}
\label{tab:diatonic-2note}
    \begin{tabular}{l c c c c}
    \toprule
    \makecell[l]{Pitch Class\\Combination} & $t$ & $I_S(X)$ &
    \makecell[c]{Instances\\in scale} & $p(x)$ \\
    \midrule
    01 & 2 & 2.58 & 2 & 2/21 \\
    02 & 5 & 1.26 & 5 & 5/21 \\
    03 & 4 & 1.58 & 4 & 4/21 \\
    04 & 3 & 2.00 & 3 & 3/21 \\
    05 & 6 & 1.00 & 6 & 6/21 \\
    06 & 2 & 2.58 & 1 & 1/21 \\
    \bottomrule
    \end{tabular}
\end{table}

%% file: tables/tab-trichords.tex
\begin{table}[htbp]
    \setlength{\tabcolsep}{4pt}
    \centering
    \caption{Information-gain values for all trichords in the diatonic set.
    Columns list: combination, information gain in bits ($I_S(X)$), probability of occurrence $p(x)$,
    and number of instances in the set.}
    \label{tab:trichords}
    \begin{tabular}{l c c c}
        \toprule
            \makecell[c]{Combination} &
            \makecell[c]{$I_S(X)$} &
            \makecell[c]{$p(x)$} &
            \makecell[c]{Instances\\in the set} \\
        \midrule
            012 & --   & 0.000 (0/35) & 0 \\
            013 & 2.58 & 0.057 (2/35) & 2 \\
            014 & --   & 0.000 (0/35) & 0 \\
            015 & 2.58 & 0.057 (2/35) & 2 \\
            016 & 3.58 & 0.029 (1/35) & 1 \\
            023 & 2.58 & 0.057 (2/35) & 2 \\
            024 & 2.00 & 0.086 (3/35) & 3 \\
            025 & 1.58 & 0.114 (4/35) & 4 \\
            026 & 3.58 & 0.029 (1/35) & 1 \\
            027 & 1.26 & 0.143 (5/35) & 5 \\
            034 & --   & 0.000 (0/35) & 0 \\
            035 & 1.58 & 0.114 (4/35) & 4 \\
            036 & 3.58 & 0.029 (1/35) & 1 \\
            037 & 2.00 & 0.086 (3/35) & 3 \\
            045 & 2.58 & 0.057 (2/35) & 2 \\
            046 & 3.58 & 0.029 (1/35) & 1 \\
            047 & 2.00 & 0.086 (3/35) & 3 \\
            048 & --   & 0.000 (0/35) & 0 \\
            056 & 3.58 & 0.029 (1/35) & 1 \\
        \bottomrule
    \end{tabular}
\end{table}

%% file: tables/tab-entropy-diatonic.tex
\begin{table}[htbp]
    \centering
    \caption{Expected information for combinations of different cardinalities in the diatonic set.}
    \label{tab:entropy-diatonic}
    \begin{tabular}{c c c c}
        \toprule
            $k$ &
            $\binom{7}{k}$ &
            $\mathbb{E}_{\Omega_k}[I]$ &
            \makecell{Candidate\\tonics} \\
        \midrule
            1 & 7  & 0.78 & 7   \\
            2 & 21 & 1.54 & 4.1 \\
            3 & 35 & 2.16 & 2.7 \\
            4 & 35 & 2.61 & 2.0 \\
            5 & 21 & 2.98 & 1.5 \\
            6 & 7  & 3.30 & 1.2 \\
            7 & 1  & 3.58 & 1.0 \\
        \bottomrule
    \end{tabular}
\end{table}

%% file: tables/tab-tonal-interpretations.tex
\begin{table}[htbp]
\setlength{\tabcolsep}{2pt} % reduce column padding
\centering
\footnotesize
\caption{Common scales and the amount of tonal ambiguity ($t$) they convey at different
combination cardinalities and within the set as a whole.}
\label{tab:tonal-interpretations}
    \begin{tabular}{l c c c c}
    \toprule
    \multicolumn{5}{c}{\textbf{Tonal Interpretations}} \\
    \midrule
    $k$ &
    \makecell[c]{Major\\(024579E)} &
    \makecell[c]{Pentatonic\\(02479)} &
    \makecell[c]{Asc.\ Mel.\\Minor\\(023579E)} &
    \makecell[c]{Harmonic\\Minor\\(023578E)} \\
    \midrule
    1 & 7.00 & 5.00 & 7.00 & 7.00 \\
    2 & 4.12 & 2.78 & 3.95 & 3.89 \\
    3 & 2.69 & 1.83 & 2.21 & 2.14 \\
    4 & 1.97 & 1.32 & 1.40 & 1.32 \\
    5 & 1.52 & 1.00 & 1.07 & 1.07 \\
    6 & 1.22 & --   & 1.00 & 1.00 \\
    7 & 1.00 & --   & 1.00 & 1.00 \\
    \midrule
    \textit{Set} & 2.42 & 2.29 & 1.92 & 1.87 \\
    \bottomrule
    \end{tabular}
\end{table}

%% file: tables/tab-tonal-interpretations-mlt.tex
\begin{table}[htbp]
\setlength{\tabcolsep}{3pt} % reduce column padding for two-column layout
\centering
\caption{Tonal interpretations ($t$) available at different note-combination cardinalities in three modes of limited transposition.}
\label{tab:tonal-interpretations-mlt}
\begin{tabular}{l c c c}
\toprule
\multicolumn{4}{c}{\textbf{Tonal Interpretations}} \\
\midrule
$k$ &
\makecell[c]{Whole tone} &
\makecell[c]{Octatonic} &
\makecell[c]{Augmented} \\
\midrule
1 & 6.00 & 8.00 & 6.00 \\
2 & 6.00 & 5.38 & 3.96 \\
3 & 6.00 & 4.42 & 3.22 \\
4 & 6.00 & 4.08 & 3.00 \\
5 & 6.00 & 4.00 & 3.00 \\
6 & 6.00 & 4.00 & 3.00 \\
7 & --   & 4.00 & --   \\
8 & --   & 4.00 & --   \\
\midrule
\textit{Set} & 6.00 & 4.33 & 3.49 \\
\bottomrule
\end{tabular}
\end{table}

%% file: tables/tab-scale-ambiguity.tex
\begin{table}[htbp]
\setlength{\tabcolsep}{4pt} % tighten column padding if needed
\centering
\caption{Popular scales and their ambiguity values. Values represent the average number of possible tonal interpretations derived from the expected-information measure in Equation~\ref{eq:diatonic-avg}.}
\label{tab:scale-ambiguity}
\begin{tabular}{l c}
\toprule
\makecell[l]{Set} & \makecell[c]{Ambiguity\\Value} \\
\midrule
Harmonic Minor (023578E)         & 1.87 \\
Asc.\ Melodic Minor (023579E)     & 1.92 \\
Major Pentatonic (02479)          & 2.29 \\
Major scale (024579E)             & 2.42 \\
Augmented (03478E)                & 3.49 \\
Octatonic (0235689E)              & 4.33 \\
Whole-tone (02468T)               & 6.00 \\
\bottomrule
\end{tabular}
\end{table}

%% file: tables/tab-comb-cardinality.tex
\begin{table}[htbp]
\setlength{\tabcolsep}{4pt} % adjust column padding for tight layouts
\centering
\caption{Note combinations of different cardinalities and their ability to identify specific 7-note sets.}
\label{tab:comb-cardinality}
\begin{tabular}{c c}
\toprule
\makecell[c]{Combination\\cardinality} &
\makecell[c]{Possible sets\\on average (range)} \\
\midrule
1 & 66 (66) \\
2 & 66 (66) \\
3 & 62.7 (39--66) \\
4 & 42.0 (14--48) \\
5 & 19.3 (13--21) \\
6 & 5.7 (1--6) \\
\bottomrule
\end{tabular}
\end{table}

%% file: tables/tab-avg-viable-sets.tex
\begin{table}[htbp]
\setlength{\tabcolsep}{4pt} % tighten column padding if needed
\centering
\caption{Average number of remaining viable common sets given combinations of different cardinalities.}
\label{tab:avg-viable-sets}
\begin{tabular}{c c}
\toprule
\makecell[c]{Combination\\cardinality} &
\makecell[c]{Possible sets\\on average (range)} \\
\midrule
1 & 6 \\
2 & 5.2 (4--6) \\
3 & 4.1 (3--5) \\
4 & 2.9 (1--5) \\
5 & 1.9 (1--4) \\
6 & 1.3 (1--2) \\
\bottomrule
\end{tabular}
\end{table}

%% file: tables/tab-appendix.tex
    \centering
    \captionsetup{type=table}
    \caption{Every Tn-class (normal form) of cardinality 1--6 that occurs in at least one of the seven reference scales. Columns give $t_S(X)$, the number of transpositions of each scale containing the combination (M = Major, mm = Asc.\ Melodic Minor, hm = Harmonic Minor, WT = Whole-Tone, O = Octatonic, P = Major Pentatonic, A = Augmented), and the number of those scales in which the combination occurs (Possible Sets). Gray cells mark single interpretations ($t_S(X)=1$); light-green rows mark combinations whose \emph{Possible Sets} equals 1.}
    \vspace{0.25\baselineskip}

\begin{center}
\begin{supertabular}{l c c c c c c c c}
% ---- cardinality 1 ----
0        & 7 & 7 & 7 & 6 & 8 & 5 & 6 & 7 \\

% ---- cardinality 2 ----
01       & 2 & 2 & 3 & -- & 4 & -- & 3 & 5 \\
02       & 5 & 5 & 3 & 6 & 4 & 3 & -- & 6 \\
03       & 4 & 4 & 5 & -- & 8 & 2 & 3 & 6 \\
04       & 3 & 4 & 4 & 6 & 4 & \sone{1} & 6 & 7 \\
05       & 6 & 4 & 4 & -- & 4 & 4 & 3 & 6 \\
06       & 2 & 4 & 4 & 6 & 8 & -- & -- & 5 \\

% ---- cardinality 3 ----
013      & 2 & 2 & 2 & -- & 4 & -- & -- & 4 \\
014      & -- & \sone{1} & 2 & -- & 4 & -- & 3 & 4 \\
015      & 2 & \sone{1} & 2 & -- & -- & -- & 3 & 4 \\
016      & \sone{1} & \sone{1} & 2 & -- & 4 & -- & -- & 4 \\
023      & 2 & 2 & 2 & -- & 4 & -- & -- & 4 \\
024      & 3 & 3 & \sone{1} & 6 & -- & \sone{1} & -- & 5 \\
025      & 4 & 3 & 2 & -- & 4 & 2 & -- & 5 \\
026      & \sone{1} & 3 & \sone{1} & 6 & 4 & -- & -- & 5 \\
027      & 5 & 3 & 2 & -- & -- & 3 & -- & 4 \\
034      & -- & \sone{1} & 2 & -- & 4 & -- & 3 & 4 \\
035      & 4 & 3 & 2 & -- & 4 & 2 & -- & 5 \\
036      & \sone{1} & 2 & 4 & -- & 8 & -- & -- & 4 \\
037      & 3 & 2 & 3 & -- & 4 & \sone{1} & 3 & 6 \\
045      & 2 & \sone{1} & 2 & -- & -- & -- & 3 & 4 \\
046      & \sone{1} & 3 & 2 & 6 & 4 & -- & -- & 5 \\
047      & 3 & 2 & 2 & -- & 4 & \sone{1} & 3 & 6 \\
048      & -- & 3 & 3 & 6 & -- & -- & 6 & 4 \\
056      & \sone{1} & \sone{1} & \sone{1} & -- & 4 & -- & -- & 4 \\

% ---- cardinality 4 ----
0134     & -- & \sone{1} & \sone{1} & -- & 4 & -- & -- & 3 \\
0135     & 2 & \sone{1} & \sone{1} & -- & -- & -- & -- & 3 \\
0136     & \sone{1} & \sone{1} & 2 & -- & 4 & -- & -- & 4 \\
0137     & \sone{1} & \sone{1} & -- & -- & 4 & -- & -- & 3 \\
0145     & -- & -- & \sone{1} & -- & -- & -- & 3 & 2 \\
0146     & -- & \sone{1} & \sone{1} & -- & 4 & -- & -- & 3 \\
0147     & -- & -- & \sone{1} & -- & 4 & -- & -- & 2 \\
0148     & -- & \sone{1} & 2 & -- & -- & -- & 3 & 3 \\
0156     & \sone{1} & -- & \sone{1} & -- & -- & -- & -- & 2 \\
0157     & \sone{1} & \sone{1} & \sone{1} & -- & -- & -- & -- & 3 \\
0158     & 2 & -- & \sone{1} & -- & -- & -- & 3 & 3 \\
\diag 0167     & -- & -- & -- & -- & 4 & -- & -- & 1 \\
0235     & 2 & 2 & \sone{1} & -- & 4 & -- & -- & 4 \\
0236     & -- & \sone{1} & \sone{1} & -- & 4 & -- & -- & 3 \\
0237     & 2 & \sone{1} & 2 & -- & -- & -- & -- & 3 \\
0245     & 2 & \sone{1} & \sone{1} & -- & -- & -- & -- & 3 \\
0246     & \sone{1} & 2 & -- & 6 & -- & -- & -- & 3 \\
0247     & 3 & 2 & -- & -- & -- & \sone{1} & -- & 3 \\
0248     & -- & 2 & \sone{1} & 6 & -- & -- & -- & 3 \\
0256     & -- & \sone{1} & -- & -- & 4 & -- & -- & 2 \\
0257     & 4 & 2 & \sone{1} & -- & -- & 2 & -- & 4 \\
0258     & \sone{1} & 2 & 2 & -- & 4 & -- & -- & 4 \\
0267     & \sone{1} & \sone{1} & \sone{1} & -- & -- & -- & -- & 3 \\
0268     & -- & 2 & -- & 6 & 4 & -- & -- & 3 \\
0346     & -- & \sone{1} & 2 & -- & 4 & -- & -- & 3 \\
0347     & -- & -- & \sone{1} & -- & 4 & -- & 3 & 3 \\
0348     & -- & \sone{1} & \sone{1} & -- & -- & -- & 3 & 3 \\
0356     & \sone{1} & \sone{1} & \sone{1} & -- & 4 & -- & -- & 4 \\
0357     & 3 & 2 & \sone{1} & -- & -- & \sone{1} & -- & 4 \\
0358     & 3 & \sone{1} & \sone{1} & -- & 4 & \sone{1} & -- & 5 \\
0367     & -- & -- & 2 & -- & 4 & -- & -- & 2 \\
0368     & \sone{1} & 2 & \sone{1} & -- & 4 & -- & -- & 4 \\
0369     & -- & -- & 4 & -- & 8 & -- & -- & 2 \\
0457     & 2 & \sone{1} & \sone{1} & -- & -- & -- & -- & 3 \\
0467     & \sone{1} & \sone{1} & \sone{1} & -- & 4 & -- & -- & 4 \\

% ---- cardinality 5 ----
01346    & -- & \sone{1} & \sone{1} & -- & 4 & -- & -- & 3 \\
\diag 01347    & -- & -- & -- & -- & 4 & -- & -- & 1 \\
01348    & -- & \sone{1} & \sone{1} & -- & -- & -- & -- & 2 \\
01356    & \sone{1} & -- & \sone{1} & -- & -- & -- & -- & 2 \\
01357    & \sone{1} & \sone{1} & -- & -- & -- & -- & -- & 2 \\
\diag 01358    & 2 & -- & -- & -- & -- & -- & -- & 1 \\
\diag 01367    & -- & -- & -- & -- & 4 & -- & -- & 1 \\
01368    & \sone{1} & \sone{1} & \sone{1} & -- & -- & -- & -- & 3 \\
01369    & -- & -- & 2 & -- & 4 & -- & -- & 2 \\
\diag 01378    & \sone{1} & -- & -- & -- & -- & -- & -- & 1 \\
\diag 01457    & -- & -- & \sone{1} & -- & -- & -- & -- & 1 \\
01458    & -- & -- & \sone{1} & -- & -- & -- & 3 & 2 \\
\diag 01467    & -- & -- & -- & -- & 4 & -- & -- & 1 \\
01468    & -- & \sone{1} & \sone{1} & -- & -- & -- & -- & 2 \\
01469    & -- & -- & \sone{1} & -- & 4 & -- & -- & 2 \\
\diag 01478    & -- & -- & \sone{1} & -- & -- & -- & -- & 1 \\
\diag 01479    & -- & -- & -- & -- & 4 & -- & -- & 1 \\
01578    & \sone{1} & -- & \sone{1} & -- & -- & -- & -- & 2 \\
02356    & -- & \sone{1} & -- & -- & 4 & -- & -- & 2 \\
02357    & 2 & \sone{1} & \sone{1} & -- & -- & -- & -- & 3 \\
02358    & \sone{1} & \sone{1} & \sone{1} & -- & 4 & -- & -- & 4 \\
\diag 02367    & -- & -- & \sone{1} & -- & -- & -- & -- & 1 \\
02368    & -- & \sone{1} & -- & -- & 4 & -- & -- & 2 \\
02369    & -- & -- & \sone{1} & -- & 4 & -- & -- & 2 \\
02457    & 2 & \sone{1} & -- & -- & -- & -- & -- & 2 \\
02458    & -- & \sone{1} & \sone{1} & -- & -- & -- & -- & 2 \\
02467    & \sone{1} & \sone{1} & -- & -- & -- & -- & -- & 2 \\
02468    & -- & \sone{1} & -- & 6 & -- & -- & -- & 2 \\
02469    & \sone{1} & 2 & -- & -- & -- & -- & -- & 2 \\
\diag 02478    & -- & \sone{1} & -- & -- & -- & -- & -- & 1 \\
02479    & 3 & \sone{1} & -- & -- & -- & \sone{1} & -- & 3 \\
02568    & -- & \sone{1} & -- & -- & 4 & -- & -- & 2 \\
02578    & \sone{1} & \sone{1} & \sone{1} & -- & -- & -- & -- & 3 \\
03467    & -- & -- & \sone{1} & -- & 4 & -- & -- & 2 \\
03468    & -- & \sone{1} & \sone{1} & -- & -- & -- & -- & 2 \\
\diag 03478    & -- & -- & -- & -- & -- & -- & 3 & 1 \\
03568    & \sone{1} & \sone{1} & -- & -- & 4 & -- & -- & 3 \\
03578    & 2 & -- & \sone{1} & -- & -- & -- & -- & 2 \\

% ---- cardinality 6 ----
\diag 013467   & -- & -- & -- & -- & 4 & -- & -- & 1 \\
013468   & -- & \sone{1} & \sone{1} & -- & -- & -- & -- & 2 \\
013469   & -- & -- & \sone{1} & -- & 4 & -- & -- & 2 \\
\diag 013479   & -- & -- & -- & -- & 4 & -- & -- & 1 \\
\diag 013568   & \sone{1} & -- & -- & -- & -- & -- & -- & 1 \\
\diag 013569   & -- & -- & \sone{1} & -- & -- & -- & -- & 1 \\
\diag 013578   & \sone{1} & -- & -- & -- & -- & -- & -- & 1 \\
\diag 013579   & -- & \sone{1} & -- & -- & -- & -- & -- & 1 \\
\diag 013679   & -- & -- & -- & -- & 4 & -- & -- & 1 \\
\diag 013689   & -- & -- & \sone{1} & -- & -- & -- & -- & 1 \\
\diag 014578   & -- & -- & \sone{1} & -- & -- & -- & -- & 1 \\
\diag 014589   & -- & -- & -- & -- & -- & -- & 3 & 1 \\
\diag 014679   & -- & -- & -- & -- & 4 & -- & -- & 1 \\
\diag 014689   & -- & -- & \sone{1} & -- & -- & -- & -- & 1 \\
023568   & -- & \sone{1} & -- & -- & 4 & -- & -- & 2 \\
\diag 023569   & -- & -- & -- & -- & 4 & -- & -- & 1 \\
023578   & \sone{1} & -- & \sone{1} & -- & -- & -- & -- & 2 \\
023579   & \sone{1} & \sone{1} & -- & -- & -- & -- & -- & 2 \\
\diag 023689   & -- & -- & -- & -- & 4 & -- & -- & 1 \\
\diag 024578   & -- & \sone{1} & -- & -- & -- & -- & -- & 1 \\
\diag 024579   & 2 & -- & -- & -- & -- & -- & -- & 1 \\
024679   & \sone{1} & \sone{1} & -- & -- & -- & -- & -- & 2 \\
\diag 024689   & -- & \sone{1} & -- & -- & -- & -- & -- & 1 \\
\diag 02468T   & -- & -- & -- & 6 & -- & -- & -- & 1 \\

\end{supertabular}
\end{center}